\documentclass[reprint,amsmath,amssymb,aps,prl,longbibliography,superscriptaddress,floatfix]{revtex4-1}

\usepackage{graphicx}
\usepackage{dcolumn}
\usepackage{bm}
\usepackage{xcolor}

\newcommand{\Rtwo}[1]{#1}

\newcommand{\affkth}{\affiliation{Department of Applied Physics, 
KTH Royal Institute of Technology, Albanova University Centre,
Roslagstullsbacken 21, 106 91 Stockholm, Sweden}}
\newcommand{\affsu}{\affiliation{Department of Physics, 
Stockholm University, 106 91 Stockholm, Sweden}}
\newcommand{\affsutd}{\affiliation{Singapore University of 
Technology and Design, 8 Somapah Road, 487372 Singapore}}
\newcommand{\affhust}{\affiliation{School of Optical and 
Electronic Information, Huazhong University of Science and 
Technology, Wuhan 430074, China}}
\newcommand{\affhustm}{\affiliation{Research Center for Cross-Scale Global-Scale Multimodal Optical Communication Discipline and Technology, Wuhan 430074, China}}
\newcommand{\affhustv}{\affiliation{Optics Valley Laboratory, 1037 Luoyu Road, Wuhan 430074, People’s Republic of China}}

\begin{document}


\title{Interferometric Readout of Momentum-Space Topology in a Programmable Dissipative Photonic Circuit}

\author{Andrea Cataldo} \email{andreacl@kth.se}\affkth
\author{Emil J. Bergholtz} \affsu
\author{Daniel Leykam} \affsutd
\author{Jun Gao} \email{jungao@hust.edu.cn}\affhust\affhustm\affhustv
\author{Ali W. Elshaari} \email{elshaari@kth.se}\affkth

\date{\today}

\begin{abstract}
Topology under non-Hermitian dynamics is encoded in the phase of the bulk evolution, yet strong dissipation suppresses the amplitudes carrying it. We resolve this using a programmable photonic integrated circuit that implements such dynamics in synthetic momentum space through unitary dilation, with the phase recovered by phase-shifted interferometry. For the non-Hermitian Su-Schrieffer-Heeger model, the method distinguishes trivial and non-trivial Zak phases and yields a coherence winding $q=\pm 1$ induced by an exceptional point. Extending to a synthetic torus via a Rice-Mele pump gives the first Chern number $\mathrm{Ch}_1=0,1$ for trivial and non-trivial cycles, showing a programmable route to momentum-space topology under strongly dissipative non-Hermitian dynamics. 
\\[1ex]
\textit{Accepted in Physical Review Letters, DOI: 10.1103/xkht-64kc.
\copyright\ 2026 American Physical Society.}
\end{abstract}

\maketitle
Topology in condensed matter and photonics is most naturally expressed in momentum-space, where invariants such as the Zak phase, Chern number, and winding numbers are defined directly on the Brillouin zone. In photonic platforms, however, these invariants are most often inferred indirectly, through their bulk-boundary correspondence as edge or interface states \cite{hafezi, noh, ozawa}. While such boundary signatures have established the field, they probe topology only through its consequences rather than its momentum-space origin, and they require finite, carefully terminated structures in which edge modes are well separated from the bulk. A complementary line of work has therefore sought to probe bulk invariants directly through the dynamics of wavepackets in real-space lattices. These dynamical approaches reach quantities ranging from Zak phases in chiral and non-chiral one-dimensional models \cite{cardano, longhi18, longhi19, wang, jiao, xu22} to non-Hermitian topological transitions \cite{zeuner,yao18,Kunst18} and topological invariants under loss \cite{leykam}. The invariant is often obtained as a bulk-averaged quantity, as in the mean chiral displacement, which converges at long times~\cite{maffei2018}. They nonetheless inherit the limitations of finite real-space implementations: discrete momentum sampling set by the lattice extent, sensitivity to finite-size and boundary effects, and limited reconfigurability once the device is fabricated. 

\begin{figure}
    \centering
    \includegraphics[width=\columnwidth]{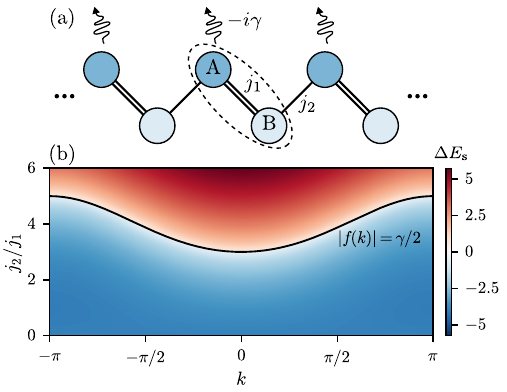}
    \caption{(a) Real-space representation of the non-Hermitian SSH lattice, with intracell and intercell couplings \(j_1\) and \(j_2\), respectively, and loss \(-i\gamma\) on the A sites. The dashed ellipse marks the \((A,B)\) unit cell. (b) Momentum-space band splitting \(\Delta E_\text{s}\), piecewise defined by \(\Delta E_\text{s}=\operatorname{Re}(E_+-E_-)\) for \(|f(k)|\ge \gamma/2\) and \(\Delta E_\text{s}=-\operatorname{Im}(E_+-E_-)\) for \(|f(k)|<\gamma/2\), shown in the \((k,j_2/j_1)\) plane for \(j_1=0.5\) and \(\gamma=4\). The black curve marks the EP condition \(|f(k)|=\gamma/2\).}
    \label{fig:nh_ssh}  
\end{figure}

\begin{figure*}
    \centering
    \includegraphics[width=\textwidth]{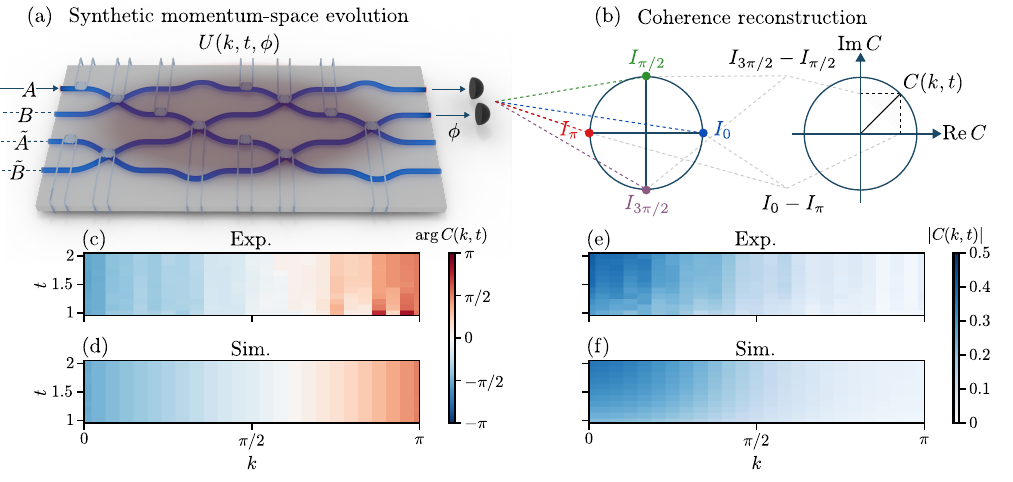}
    \caption{(a) Schematic of the interferometric readout for synthetic momentum-space evolution on a programmable PIC. Light is launched into spatial mode \(A\), with \((A,B)\) spanning the signal subspace and \((\tilde{A},\tilde{B})\) the ancilla subspace. The PIC implements a compiled \(4\times4\) transformation \(U(k,t,\phi)=U_\text{int}(\phi)U_\text{dil}(k,t)\), which combines the dilation and phase-shifted interferometric readout. Two quadratures are formed as shown in (b), allowing reconstruction of the coherence \(C(k,t)\) as in Eq.~\eqref{eq:C}. For the parameters \(j_1=0.5\), \(j_2=1.5\), and \(\gamma=4\), the relative phase \(\operatorname{arg}C(k,t)\) is obtained for the experiment (c) and simulation (d), with (e,f) showing the corresponding magnitude \(|C(k,t)|\). All data are shown for a representative well-conditioned interval \(t\in [1,2]\), with \(N_k=25\) uniformly spaced momentum points over \(k\in[0,\pi]\). The root-mean-square errors (RMSEs) relative to simulation are \(0.0613\pi\) for the phase and \(0.0194\) for the magnitude.}
    \label{fig:readout}
\end{figure*}

In parallel, grating-based, metasurface, and photonic-crystal systems have provided more direct momentum-space access, with topological invariants extracted from reflection or far-field radiation in one-dimensional photonic crystals and metasurfaces \cite{gao, gorlach, liu} and from non-Hermitian guided-mode resonances at nanophotonic interfaces \cite{lee}. These approaches yield clean spectroscopic signatures but are tied to fixed geometries, so that scanning across the parameter space of a topological model requires fabricating a new sample. These considerations motivate a complementary route that combines the direct momentum-space access of these spectroscopic platforms with the reconfigurability needed to traverse topological phase diagrams, and that remains viable in the strongly dissipative regime where non-Hermitian physics is richest \cite{eg, bergholtz}. Strong dissipation is particularly demanding because it suppresses the evolved amplitudes whose relative phase carries the topological information.

Here, a programmable photonic integrated circuit (PIC) \cite{bogaerts, on} is used to simulate non-Hermitian dynamics directly in synthetic momentum-space, with each momentum point realized as a separately programmed unitary on the same hardware. Such pointwise access differs from a synthetic dimension formed by dynamically coupled modes~\cite{lin2022}. The method requires neither a real-space boundary nor spectrally resolved edge modes. The experimental implementation builds on the platform of Ref.~\cite{xu25}. Strong dissipation is handled by embedding the non-unitary evolution into a larger unitary via Sz.-Nagy dilation \cite{hu2020} --- an approach also used to realize non-Hermitian dynamics on spin and superconducting platforms \cite{wu2019,dogra2021} --- while the phase of the evolved signal is recovered through phase-shifted interferometric readout. The method is first demonstrated in the non-Hermitian SSH model, where it extracts the Zak phase and reveals an interferometric signature of an exceptional point, before being extended to a higher-dimensional synthetic space through the first Chern number of a non-Hermitian Rice-Mele pump.

The model is shown in Fig.~\ref{fig:nh_ssh}(a), and its momentum-space Hamiltonian reads
\begin{equation}
    H(k) =
    \begin{bmatrix}
        -i\gamma & f(k) \\
        f^{*}(k) & 0
    \end{bmatrix},
    \qquad 
    f(k) = j_1 + j_2 e^{-ik},
\label{eq:H}
\end{equation}
where \(j_1\) and \(j_2\) denote the intracell and intercell couplings, respectively. The coupling function \(f(k)\) traces a trajectory in the \((\operatorname{Re} f, \operatorname{Im} f)\) plane that determines the winding and hence the topology. A loss term \(-i\gamma\) is included directly in \(H(k)\), so that for \(|f(k)|<\gamma/2\) the dynamics lies in the dissipative regime; there, the two bands are distinguished not by \(\operatorname{Re} E_{\pm}(k)\) but by \(\operatorname{Im} E_{\pm}(k)\), i.e. by their decay rates, as shown in Fig.~\ref{fig:nh_ssh}(b).

For each \(k\), the evolution over time \(t\) is described by the propagator \(e^{-iH(k)t}\). Because the PIC natively implements unitary transformations, this propagator is rescaled by its largest singular value so as to become a contraction, and is thereby dilated to a Sz.-Nagy unitary \(U_\text{dil}(k,t)\). The target evolution then acts on the \((A,B)\) signal subspace of the SSH model, while \((\tilde{A},\tilde{B})\) provide the ancilla subspace that makes the unitary realization possible, as shown in Fig.~\ref{fig:readout}(a) and detailed in the Supplemental Material~\cite{SM}.

\begin{figure*}
    \centering
    \includegraphics[width=\textwidth]{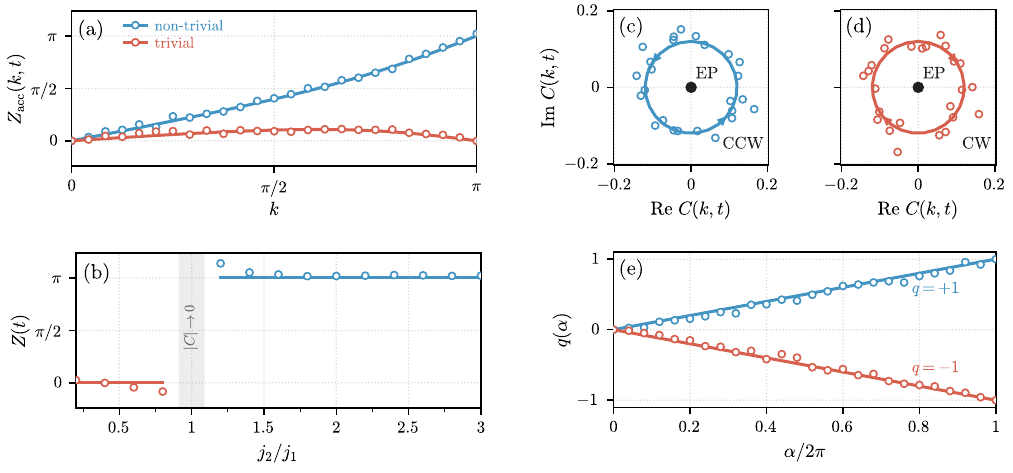}
    \caption{Extraction of momentum-space topology. (a) Accumulated Zak phase \(Z_\text{acc}(k,t)\), as defined by Eq. \eqref{eq:Z_acc}, for the non-trivial and trivial cases, corresponding to \((j_1,j_2,\gamma)=(0.5, 1.5, 4)\) and \((1.5, 0.5, 4)\), respectively. Circles denote experiment and lines the simulation. The extracted values at \(k=\pi\) are \(1.02 \pi\) and \(-0.00148\pi\). (b) Zak phase \(Z(t)\) versus \(j_2/j_1\) for fixed \(j_1=0.5\), showing quantization to \(0\) and \(\pi\), with average values of \(-0.0254\pi\) and \(1.04\pi\) on the two sides of the transition. A shaded region near the transition indicates \(|C|\to 0\). Panels (a,b) are shown for \(t=2\), with \(N_k=25\) uniformly spaced momentum points over \(k\in[0,\pi]\), i.e. 100 unitaries per parameter set. (c,d) Encirclement of an EP-induced coherence zero using the same circular contour with reversed traversal, counter-clockwise (CCW) and clockwise (CW). Circles denote the measured \(C(k,t)\) and lines the simulation. (e) EP-induced coherence winding \(q(\alpha)\) versus \(\alpha/2\pi\), obtained by integrating according to Eq. \eqref{eq:q}. Both simulation and experiment give \(q=1\) for CCW and \(q=-1\) for CW.}
    \label{fig:zak}
\end{figure*}

\begin{figure*}
    \centering
    \includegraphics[width=\textwidth]{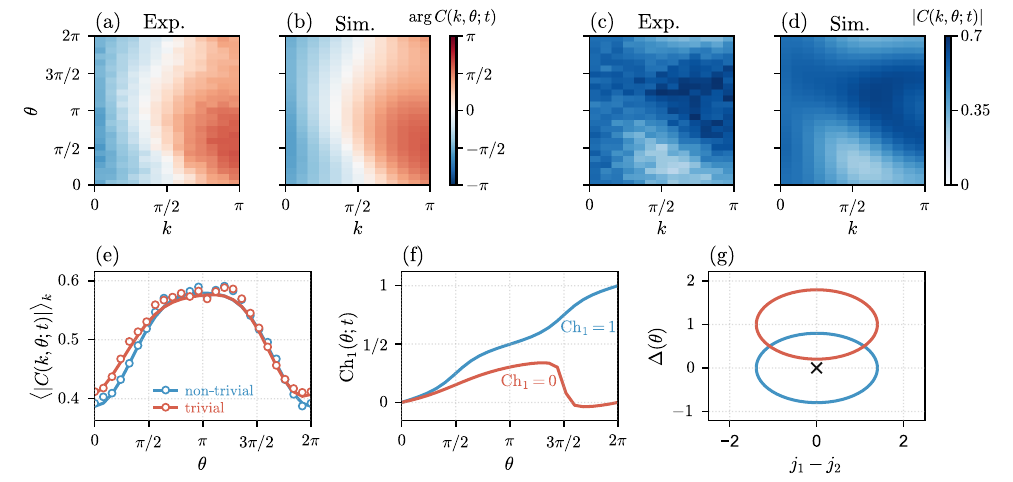}
    \caption{Higher-dimensional synthetic space extension of a non-Hermitian Rice-Mele pump. (a-d) Reconstructed coherence of the non-trivial pump cycle with \((j_0,\delta,\Delta_c,\Delta_0,\gamma,t)=(2,0.7,0,0.8,2,0.25)\), shown as phase and magnitude for experiment and simulation. The input state is \((1,-i)^T/\sqrt{2}\), giving RMSEs relative to simulation of \(0.0208\pi\) and \(0.0302\), respectively. (e) Momentum-averaged magnitude \(\langle |C(k,\theta;t)| \rangle_k\) for non-trivial and trivial pump cycles. Lines denote simulation and circles experiment; the RMSEs of the averaged traces are \(0.00769\) and \(0.00864\), respectively. (f) Accumulated first Chern number inferred from the experimentally reconstructed pump trajectory, giving \(\operatorname{Ch}_1 = 1\) and \(\operatorname{Ch}_1 = 0\) for the non-trivial and trivial cycles, respectively. (g) Pump cycles in the \((j_1-j_2,\Delta)\) plane relative to the gap closing (cross), with the trivial cycle shifted upward by \(\Delta_c=1\) so that only the non-trivial cycle encloses it. The fitted cycles preserve this separation, with margins \(0.805\) (non-trivial) and \(0.175\) (trivial) from the gap closing, which keeps the Chern number extraction in (f) stable. All data use \(N_k=N_\theta=25\) uniformly spaced points over \(k,\theta\in[0,2\pi)\), i.e. 2500 unitaries per cycle; panels (a--d) display only \(k\in[0,\pi]\), over which their quoted RMSEs are evaluated.}
    \label{fig:rm}
\end{figure*}

The evolved \((A, B)\) signal amplitudes are then subjected to interferometric readout through \(U_\text{int}(\phi)\), which mixes them against four phase-shifted references at \(\phi=0\), \(\pi/2\), \(\pi\), and \(3\pi/2\). These readings provide the phase quadratures \cite{psi}, shown in Fig.~\ref{fig:readout}(b), from which one forms the interferometric coherence
\begin{equation}
    C(k,t) = \big[ I_{0} - I_{\pi} \big]
    +  i\big[ I_{3\pi/2} - I_{\pi/2} \big],
\label{eq:C}
\end{equation}
where \(I_{\phi}\) denotes the output intensity at phase reference \(\phi\), with the \((k,t)\) dependence omitted for brevity. The phase \(\operatorname{arg}C(k,t)\) gives the relative phase of the \(A\) and \(B\) amplitudes, while \(|C(k,t)|\) sets the strength of the readout.  

An example of this interferometric readout is shown in Fig.~\ref{fig:readout}(c-f) for the non-trivial case with \(j_1=0.5\), \(j_2=1.5\), and \(\gamma=4\). The measured phase readout reproduces the main features of the corresponding simulation. Deviations occur mainly in regions of diminished coherence. This is because as \(|C(k,t)|\) becomes small, the phase becomes increasingly ill-conditioned. In such cases this may produce abrupt phase variations, but away from these regions the phase varies smoothly. 

These deviations should not be confused with generic disorder in a real-space lattice, which would break translational symmetry and couple different momenta. Indeed, the finite fidelity of the programmed unitary is the relevant experimental imperfection. It can be mitigated by using a calibrated model of an imperfect PIC \cite{fyrillas2024, zheng2024, Fan} that accounts for the actual device response~\cite{SM}.

By repeating the readout at successive \(k\)-points, one traces the phase across the BZ. For the chosen input state, \(\partial_k \arg C=\partial_k \arg f^*(k)\) \cite{SM}. This in turn provides access to the Zak phase, \Rtwo{here that of the Hermitian parent model}, with its accumulation defined by
\begin{equation}
    Z_\text{acc}(k,t) = \int_{0}^{k} dk' \, \partial_{k'} \operatorname{arg} C(k',t).
\label{eq:Z_acc}
\end{equation}
Fig.~\ref{fig:zak}(a) shows \(Z_\text{acc}(k,t)\) for the non-trivial and trivial cases. In the non-trivial case, it increases progressively with \(k\) and reaches nearly \(\pi\) at the boundary, consistent with a net winding. In the trivial case, it shows only a small excursion and returns nearly to zero at the end, indicating vanishing net winding. This behavior is consistent with the underlying model, where \(f(k)\) encloses the origin in the former case, whereas it does not in the latter. The distinction of topology is thus already evident throughout the accumulated phase trace. The same distinction is seen in Fig.~\ref{fig:zak}(b), which plots the Zak phase \(Z(t)=Z_\text{acc}(\pi,t)\) against \(j_2/j_1\) for fixed \(j_1=0.5\). There, the measured values lie near \(0\) for \(j_2/j_1<1\) and near \(\pi\) for \(j_2/j_1>1\).

One may also probe the interferometric signature of the EP itself. The accumulated phase is evaluated along a closed contour in the \((\operatorname{Re}C,\operatorname{Im}C)\) plane. An EP-induced coherence winding can be obtained by
\begin{equation}
    q(\alpha) = \frac{1}{2\pi} \int_{\mathcal{L}(\alpha)} d \, \operatorname{arg} C(k,t),
\label{eq:q}
\end{equation} \
where \(\mathcal{L}(\alpha)\) is this contour traced up to \(\alpha\). In the coherence plane, the EP manifests itself as the point \(C(k,t)=0\). Therefore, a convenient choice is to take the contour as a circle within this plane, so that \(|C(k,t)|\) is kept finite. This is visualized in Fig.~\ref{fig:zak}(c,d) for two orientations of the same contour. The resulting \(q(\alpha)\) progresses linearly, with opposite windings \(q = \pm 1\), as seen in Fig.~\ref{fig:zak}(e). We emphasize that $q$ as defined in Eq.~\eqref{eq:q} is distinct from the conventional eigenvalue vorticity associated with a second-order EP, which arises from the branch-point structure and yields a half-integer winding $\pm 1/2$ upon encircling the EP in parameter space. By contrast, $q$ 
counts the integer winding of $\arg C(k,t)$ around the zero of the interferometric coherence induced by the EP. This provides a complementary characterization of the EP, based on the phase imprinted on the experimentally accessible coherence signal rather than on direct observation of the eigenvalue spectrum. 

As an extension to higher-dimensional synthetic space,  consider the non-Hermitian Rice-Mele model with a pump parameter \(\theta \in [0,2\pi)\), so that \((k,\theta)\) spans a synthetic torus. The Hamiltonian reads
\begin{equation}
    H_{\text{RM}}(k,\theta) =
    \begin{bmatrix}
        \Delta(\theta)-i\gamma & f(k,\theta) \\
        f^{*}(k,\theta) & -\Delta(\theta)
    \end{bmatrix},
\label{eq:H_RM}
\end{equation}
where \(f(k,\theta) = j_1(\theta) + j_2(\theta)e^{-ik}\), \(j_{1,2}(\theta)=j_0 \pm \delta \cos \theta\), and \(\Delta(\theta)=\Delta_c + \Delta_0 \sin \theta\). Over one pump cycle, the associated first Chern number can then be obtained from the winding of the Zak phase as \cite{Thoulesspump}
\begin{equation}
    \operatorname{Ch}_1 = \frac{1}{2\pi} \int_0^{2\pi} d\theta \, \partial_\theta Z(\theta).
\label{eq:Ch1}
\end{equation}

The coherence now becomes \(C(k,\theta;t)\) and is determined using the same interferometric method, with its phase and magnitude shown for non-trivial pumping in Fig.~\ref{fig:rm}(a-d). The experiment follows closely with the simulation; indeed, the momentum-averaged magnitude \(\langle |C(k,\theta;t)| \rangle_k\) stays clearly resolved for both cycles, well above the low-coherence regime, as seen in Fig.~\ref{fig:rm}(e). 

This measured coherence is then used to infer the pump trajectory. A candidate Rice-Mele loop is specified in the \((j_1-j_2, \Delta)\) plane, from which the expected coherence is computed and compared with the measured one. A fit of these quantities gives the invariant shown in Fig.~\ref{fig:rm}(f), taking the values \(\operatorname{Ch}_1=0,1\) for the trivial and non-trivial cycles, respectively. This is consistent with whether the corresponding Rice-Mele loop misses or encloses the gap closing in Fig.~\ref{fig:rm}(g). For the chosen parameters, these fitted loops remain well separated from the gap closing, so small readout errors only deform the pump trajectories without changing their winding.

In conclusion, we have demonstrated direct interferometric readout of non-Hermitian topological invariants in synthetic momentum space, resolving the central experimental challenge that strong dissipation suppresses the amplitudes carrying the topological phase. The method proceeds by (i) implementing the dilated evolution, (ii) reconstructing the coherence from four phase-shifted intensity measurements, and (iii) extracting the corresponding topological invariant from this coherence. From this readout, the Zak phase and an EP-induced coherence winding of the SSH model are obtained, together with the first Chern number of a Rice-Mele pump, thereby demonstrating a programmable route to higher-dimensional momentum-space topology in photonics. An \(N\)-band model requires \(N\) signal and \(N\) ancillary modes~\cite{SM}. The reconfigurability of the platform makes it naturally suited to the experimental investigation of non-Hermitian symmetry classes and topological invariants that remain inaccessible to fixed-geometry platforms.

\subsection*{Acknowledgments}
A.C and A.W.E acknowledge the support from Knut and Alice Wallenberg (KAW) Foundation through the Wallenberg Centre for Quantum Technology (WACQT). E.J.B. acknowledges the support from Knut and Alice Wallenberg Foundation (2023.0256) and the Göran Gustafsson Foundation for Research in Natural Sciences and Medicine. D. L. acknowledges support from the Ministry of Education, Singapore, under its SUTD Kickstarter Initiative (Grant No. SKI 20210501). J.G. acknowledges support from Swedish Research Council (Ref: 2023-06671 and 2023-05288), Vinnova project (Ref: 2024-00466) and the Göran Gustafsson Foundation. A.W.E acknowledges support from Swedish Research Council (VR) Starting Grant (Ref: 2016-03905), and Vinnova quantum kick-start project 2021.

\bibliography{references}

\end{document}